\documentclass[twocolumn,showpacs,prl]{revtex4}
\usepackage{graphicx}%
\usepackage{dcolumn}
\usepackage{amsmath}
\begin{document}

\preprint{HEP/123-qed}

\title{Valley splitting of AlAs two-dimensional electrons in a perpendicular magnetic field}
\author{Y. P. Shkolnikov, E. P. De Poortere, E. Tutuc, and M. Shayegan}
\affiliation{%
Department of Electrical Engineering, Princeton University, Princeton, New Jersey 08544
}%

\date{\today}

\begin{abstract}
By measuring the angles at which the Landau levels overlap in tilted magnetic fields (the coincidence method), we
determine the splitting of the conduction-band valleys in high-mobility two-dimensional (2D) electrons confined to
AlAs quantum wells. The data reveal that, while the valleys are nearly degenerate in the absence of magnetic
field, they split as a function of perpendicular magnetic field.  The splitting appears to depend primarily on the
magnitude of the perpendicular component of the magnetic field, suggesting electron-electron interaction as its
origin.
\end{abstract}

\pacs{73.21, 73.23, 73.43}

\maketitle

In many semiconductors, the energy dispersion of the conduction
band contains more than one minimum, or valley. Examples include
Si and AlAs, where electrons occupy pockets near or at the
equivalent X-points of the Brillouin zone. One of the
long-standing and controversial problems in the physics of
two-dimensional electron systems (2DESs) in such semiconductors
has been the nature of valley splitting, i.e., the lifting of this
valley degeneracy. While it is clear that mechanisms that break
the symmetry of the crystal potential, such as uniaxial strain,
can lift the degeneracy, it has remained controversial whether
interaction between the electrons can also lead to a splitting
\cite{Ando1982}. In this Letter, we report measurements of
transport in low disorder 2DESs in modulation-doped AlAs quantum
wells, which provide clear evidence for the dependence of valley
splitting on the applied perpendicular magnetic field ($B_\perp$).
The data reveal that at zero magnetic field the electrons occupy
two nearly-degenerate in-plane conduction-band valleys. With the
application of $B_\perp$, we observe a splitting of the valley
energies that increases monotonically with $B_\perp$ and is
essentially independent of the parallel component of the magnetic
field ($B_{||}$). This suggests that electron-electron interaction
is responsible for the splitting.

We performed experiments on 2DESs confined to modulation doped
AlAs quantum wells grown by molecular beam epitaxy on (100) GaAs
substrates. In these samples, the AlAs quantum well is flanked by
undoped and Si-doped layers of Al$_{0.4}$Ga$_{0.6}$As
\cite{Etienne2002}.  We studied three samples with two different
quantum well widths: $150\AA$ for S1 and S3, and $110\AA$ for S2.
In bulk AlAs, electrons occupy three valleys that are located at
X-point (rather than at $\Delta$-point as in Si) of the Brillouin
zone. Previous studies \cite{Yamada1992, Stergios1999, Lay1993,
Smith1988} have indicated that strain due to lattice mismatch
between AlAs quantum well and GaAs substrate leads to a lifting in
energy and depopulation of the out-of-plane valley in AlAs quantum
wells wider than 60\AA. Consistent with these studies, the 2DES in
our samples occupies two valleys whose principal axes lie in the
2D plane. In these remaining valleys, the cyclotron electron
effective mass $m^*$ is 0.46 (in units of the free electron mass,
$m_0$) and the band g-factor $g_b$ is 2. We measured the
magnetoresistance of the samples in L-shaped Hall bars aligned
with [001] and [010] directions. Using illumination and front/back
gate biasing, we were able to vary the 2D electron density $n$
between $4.0\times10^{11}$ and $9.6\times10^{11}$ cm$^{-2}$.  At
30mK, the electron mobility of our samples was as high as 250,000
cm$^2$/Vs.

To measure the valley splitting ($\Delta E_V$) as a function of
magnetic field, we utilize the coincidence method \cite{Fang1968}.
This technique makes use of the difference in the magnetic field
dependencies of the orbital and spin energies, and has been widely
used to measure the spin-splitting and the effective g-factor in
2D systems. In a 2D system in a strong magnetic field,
quantization of the orbital motion leads to the formation of
Landau levels (LLs). The energy separation between the LLs is
equal to the cyclotron energy, which, in an ideal 2D system equals
$\hbar e B_\perp/(m^*m_0)$ and therefore depends only on
$B_\perp$. Thanks to the Zeeman coupling, each LL splits into two
energy levels, one for each polarization of spin. The energy
separation between these levels, the Zeeman energy, is generally a
function of the total magnetic field ($B_{tot}$) and is equal to
$|g^*|\mu_B B_{tot}$ where $\mu_B$ is Bohr magneton and $g^*$ is
the effective g-factor.  In a typical coincidence measurement, the
2D sample is placed in a magnetic field whose direction makes an
angle $\theta$ with the normal to the 2D plane.  The Zeeman energy
is then measured (in units of the cyclotron energy) from the
values of $\theta$ at which every other magnetoresistance minima
disappear; the disappearance is the result of the overlap, at the
Fermi energy, between two energy levels associated with opposite
spin. Now, in a two-valley system, yet another splitting appears:
each spin-resolved level splits in two levels, separated in energy
by $\Delta E_V$, corresponding to each of the valleys. In our
measurements presented here, by carefully monitoring $\theta$ at
which magnetoresistance minima disappear, we are able to determine
both the Zeeman- and valley-splittings.

We first present data to establish that, at low magnetic fields,
the 2D electrons in our samples occupy two nearly-degenerate
valleys. Magnetoresistivity $\rho_{xx}$ as a function of $B_\perp$
(at zero $B_{||}$) for sample S1 at a density of
$n=9.2\times10^{11}$ cm$^{-2}$ is shown in Fig. 1 \footnote{We
determine the total 2D electron density from either the Hall
coefficient or the positions of quantum Hall states which are
observed at higher $B_\perp$.}. Strong minima in $\rho_{xx}$ occur
at every fourth Landau level filling factor $\nu$, which indicates
a fourfold degeneracy of the Landau levels (two for spin and two
for valley) \footnote{As will become clear later in the paper,
whether the strongest minima occur at $\nu=8$, 12, 16, etc., or at
$\nu=10$, 14, 20, etc. depends on the relative strengths of the
cyclotron and Zeeman energies}.  This degeneracy can be clearly
seen in the Fourier transform of $\rho_{xx}$ vs. $1/B_\perp$ data,
shown in the inset to Fig. 2 for sample S2 at a density of
$7.9\times 10^{11}$ cm$^{-2}$.  The frequency of a peak in the
Fourier transform can be converted to a 2D density by multiplying
by $e/h$, where $e$ is the electron charge and $h$ is the Planck
constant.  The presence of a peak (square symbol) at the frequency
equal to one quarter of the frequency associated with the total
density confirms the spin- and valley-degeneracy at low $B_\perp$.
We assign the other peaks in the Fourier transform to the spin-
and/or valley-resolved oscillations which are observed at higher
$B_\perp$ \footnote{The peak near zero frequency is an artifact of
the window and the range used in taking the Fourier transform.}.

In Fig. 2 we show a plot of the measured peak positions in Fourier
transforms as we change the density in S2 via a front-gate bias.
The valley/spin degeneracies persist at all $n$ and, as the
density is decreased, all the peak positions linearly decrease and
extrapolate to approximately zero in the $n=0$ limit.  It is
particularly noteworthy that the low-frequency peak does not show
any splitting within the resolution of the Fourier transform at
any density; this observation indicates that the densities of
states and therefore the effective masses of electrons in the two
occupied valleys are the same, consistent with the 2DES occupying
two nearly degenerate valleys whose principal axes lie in the 2D
plane.
\begin{figure}
\includegraphics[scale=.9]{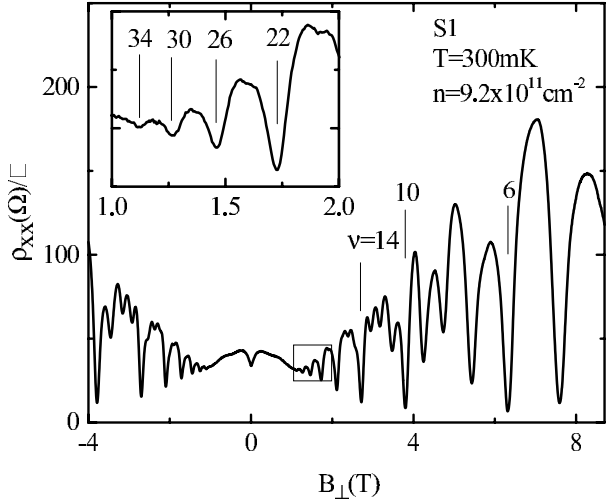}
\caption{Magnetoresistivity $\rho_{xx}$ as a function of perpendicular magnetic field $B_\perp$ for AlAs 2D
electrons in sample S1. At low $B_\perp$, $\rho_{xx}$ in the inset shows strong minima at every fourth filling
factor $\nu$, indicating a fourfold degeneracy of the Landau levels (2 for spin and 2 for valley). At higher
$B_\perp$, all minima become stronger than their ($\nu+4$) counterpart, implying that both spin- and
valley-splitting increase with $B_\perp$.} \label{real space}
\end{figure}

\begin{figure}
\includegraphics[scale=.9]{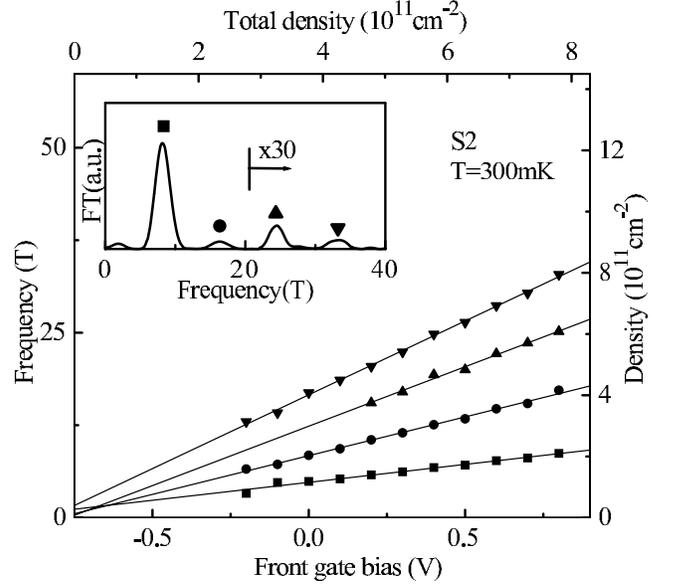}
\caption{Inset: An example of the Fourier transform of $\rho_{xx}$
vs. $1/B_\perp$ for sample S2 at $n=7.9\times10^{11}$ cm$^{-2}$.
The presence of a peak at the frequency equal to one quarter of
frequency associated with the total density confirms the
degeneracy of the spin and valleys at low $B_\perp$. Main: Fourier
transform peak frequencies as a function of the front-gate bias.}
\label{fg}
\end{figure}

Figure 1 reveals that, at high $B_\perp$, all the $\rho_{xx}$
minima become stronger than their ($\nu+4$) counterparts.  For
example, the $\rho_{xx}$ minimum at $\nu=9$ is deeper than the
minimum at $\nu=13$, while at $\nu=17$ there is no visible
minimum.  This progression suggests that all the three relevant
energies in the system, i.e., the cyclotron energy, Zeeman energy,
and $\Delta E_V$, increase with increasing $B_\perp$.  Our
coincidence measurements, summarized in Fig. 3(a), provide for a
quantitive determination of the energies.  This figure, which is
the highlight of our study, shows a (color) plot of $\rho_{xx}$
vs. $1/cos \theta$ (x-axis) and filling factor, $\nu$ (y-axis).
The plot was made by taking magnetoresistance traces at 32
different $\theta$'s. A striking alternating diamond pattern
emerges in the data. This pattern also has periodicity of four in
$\nu$. Surprisingly, the pattern changes very little over the
($B$, $\theta$) parameter space; as shown later, this observation
implies that $\Delta E_V$ increases approximately linearly with
$B_\perp$. Such a dependence leads to the simple energy
fan-diagram shown in Fig. 3(b).  Another prominent feature of the
pattern is its right-left symmetry, which indicates that $\Delta
E_V$ is independent of $B_{||}$.

In order to make a quantitative analysis of the data, we assume a general model for the energies of a two-valley
system tilted in a magnetic field:
\begin{equation}\label{energy levels v1}
  E_{1}=(N+\frac{1}{2})\frac{\hbar e B_\perp}{m_0 m^*}\pm\frac{1}{2}|g^*|\mu_B
  \frac{B_\perp}{cos\theta}-\frac{\Delta E_V}{2}
\end{equation}
\begin{equation}\label{energy levels v2}
    E_{2}=(N'+\frac{1}{2})\frac{\hbar e B_\perp}{m_0 m^*}\pm\frac{1}{2}|g^*|\mu_B
  \frac{B_\perp}{cos\theta}+\frac{\Delta E_V}{2},
\end{equation}
where $E_{1}$ and $E_{2}$ are the energy levels of valleys 1 and
2, respectively, $N$ and $N'$ are the LL indices, and $g^*$ is the
effective g-factor \footnote{As we show later, $g^*$ is the same
for both valleys.}. By fitting the coincidences between energy
levels of the same valley using this model, we can determine the
Zeeman energy (in units of the cyclotron energy), or equivalently
the product $|g^*|m^*$, as a function of $B_\perp$ and $B_{||}$.
Likewise, from the coincidences between the energy levels of the
different valleys, we can determine $\Delta E_V$ (in units of the
cyclotron energy), or equivalently, the product $m^*\Delta E_V$.
\begin{figure}
\includegraphics[scale=.9]{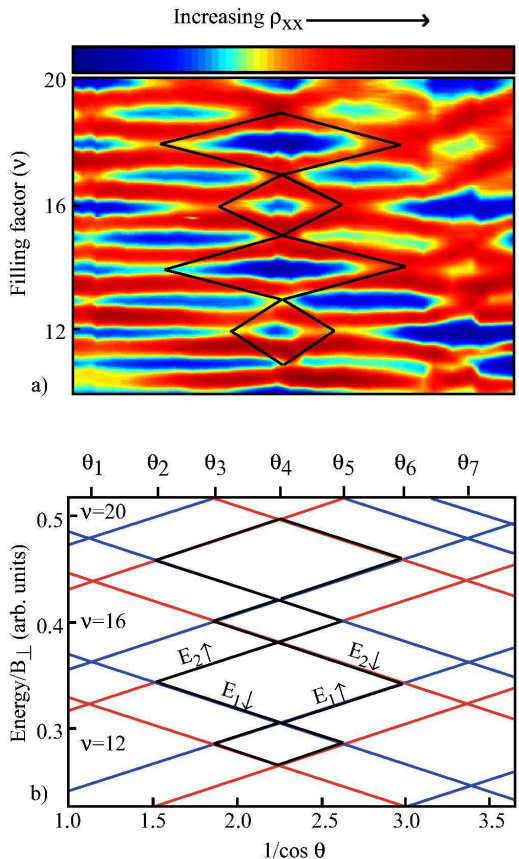}
\caption{ a) Magnetoresistivity $\rho_{xx}$ vs. $1/cos\theta$ and $\nu$ measured in sample S3 at T=30mK. Blue and
red colors represent small and large values of $\rho_{xx}$, respectively. To increase the visibility of
$\rho_{xx}$ oscillations over the whole field range, we scaled $\rho_{xx}$ by a Dingle-like factor, $exp(1/6.4
B_\perp)$. A striking alternating diamond pattern emerges in the data. This pattern has an approximate right-left
symmetry, has a periodicity of four in $\nu$, and changes little over the whole field range. b) Fan diagram that
qualitatively accounts for the data. In this diagram, $\Delta E_V$ scales linearly with $B_\perp$ and is
independent of $B_{||}$.} \label{autonum}
\end{figure}

We first determine the product $|g^*|m^*$ directly from
coincidence angles $\theta_1$, $\theta_4$, and $\theta_7$ in Fig.
3(a); these angles correspond to the crossings between the LLs
associated with the same valley. We find that, at low fields,
$|g^*|m^*$ is constant, identical for both valleys, and at
$n=9.55\times10^{11}$ cm$^{-2}$ equal to $1.76\pm0.02$.
Surprisingly, while $|g^*|m^*$ is enhanced with respect to its
band value of 0.92, it does not oscillate as function of $\nu$
contrary to the theoretical expectation \cite{Ando1974}. Papadakis
\textit{et al.} \cite{Stergios1999} have also observed a constant
enhancement of $|g^*|m^*$ in low density ($n=2\times10^{11}$
cm$^{-2}$) AlAs 2D electrons for $\nu\geq9$. In the higher density
samples that we have studied, however, we find a monotonic
increase in $|g^*|m^*$ for $\nu<7$. Here we limit our
determination of $\Delta E_V$ to the field range where the
measured $|g^*|m^*$ is constant.

Next, we determine $\Delta E_V$ from the values of angles
$\theta_2, \theta_3, \theta_5$ and $\theta_6$ at which the energy
levels of different valleys overlap.  Since we do not know exactly
the absolute LL index, but only the difference between the indices
of LLs to which these energy levels belong, $\Delta E_V$ can be
determined only to within an additive term $2l\hbar e
B_\perp/m^*m_0$, where $l$ is an integer. In sample S2, the
presence of coincidences at $\theta_2$ for low $\nu$ (not shown)
requires $l$ to be zero. For sample S3, coincidences indicate that
the system is not valley-polarized for $\nu\geq8$. This
observation restricts $l$ to either zero or one. We chose zero,
since $l=1$ implies the unlikely situation that valley splitting
first shrinks to zero and then increases as a function of
$B_\perp$.

Valley splitting measured for samples S2 and S3 is summarized in
Fig. 4. Since $\Delta E_V$ does not vary with the coincidence
angle, it is independent of $B_{||}$ \footnote{For example the two
data points at $B_\perp$=2.82T are from coincidences at
$\theta=47^o$ and $\theta=71^o$}. Furthermore, $\Delta E_V$ for
different samples and densities fall on the same curve which
exhibits a nearly linear dependence on $B_\perp$ for
$B_\perp>1.5$T.  A least squares fit of the data in this
high-field range gives $\Delta E_V=-0.22+0.25B_\perp$ (assuming
$m^*=0.46$), with $\Delta E_V$ in units of meV and $B_\perp$ in
units of Tesla. At low fields, $\Delta E_V$ deviates from the
linear behavior.

Two features of the data in Fig. 4 are noteworthy.  First, since
the widths of the LLs do not affect the positions of the
coincidences, there should be no disorder corrections to the
plotted values of $\Delta E_V$ \footnote{The FWHM of the electron
wavefunction in our samples is likely less than 50 \AA, so that
corrections for the effects of the finite layer thickness are less
than 1\% of $\Delta E_V$.}. Second, the fact that $\Delta E_V$
appears to be independent of density and $B_{||}$, and depends
only on $B_\perp$, suggests that electron-electron interaction is
responsible for its enhancement with $B_\perp$. The linear
dependence on $B_\perp$, on the other hand, is puzzling; one would
normally expect a $B^{1/2}$ dependence as the enhancement should
inversely scale with the magnetic length.  We hope that the
results presented in this Letter will serve as incentive for
developing a theory to explain the enhancement of valley-splitting
in a magnetic field.

\begin{figure}
\includegraphics[scale=.9]{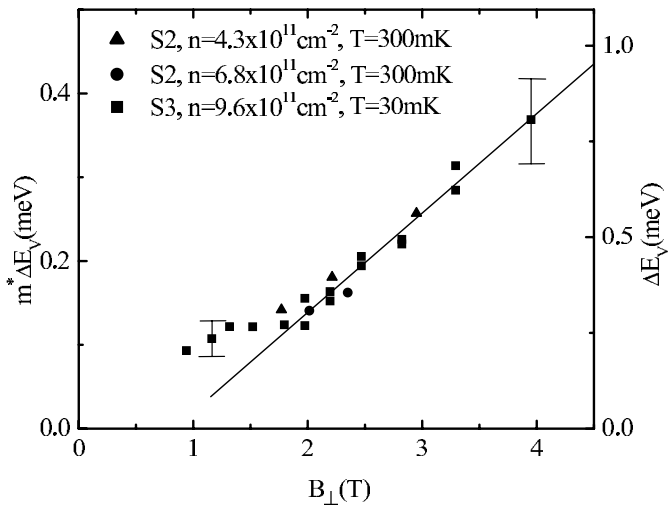}
\caption{Products $m^*\Delta E_V$ and $\Delta E_V$ (assuming
$m^*$=0.46) as a function of $B_\perp$ with typical error bars.
Solid line is the least squares linear fit for the high field
valley splitting.} \label{dE_V}
\end{figure}

\begin{acknowledgments}
We acknowledge the NSF support for this work.  Part of the
measurements were performed at the NSF-supported National High
Magnetic Field Laboratory in Tallahassee, Florida; we thank E.
Palm and T. Murphy for technical assistance. We would also like to
thank D. Tsui for helpful discussions.
\end{acknowledgments}

\bibliography{valley_final}
\end{document}